# Blind Compressive Sensing Framework for Collaborative Filtering

Anupriya Gogna
ECE Department
IIIT-Delhi
Delhi, INDIA
anupriyag@iiitd.ac.in

Angshul Majumdar
ECE Department
IIIT-Delhi
Delhi, INDIA
angshul@iiitd.ac.in

## ABSTRACT
Existing works based on latent factor models have focused on representing the rating matrix as a product of user and item latent factor matrices, both being dense. Latent (factor) vectors define the degree to which a trait is possessed by an item or the affinity of user towards that trait. A dense user matrix is a reasonable assumption as each user will like/dislike a trait to certain extent. However, any item will possess only a few of the attributes and never all. Hence, the item matrix should ideally have a sparse structure rather than a dense one as formulated in earlier works. Therefore we propose to factor the ratings matrix into a dense user matrix and a sparse item matrix which leads us to the Blind Compressed Sensing (BCS) framework. We derive an efficient algorithm for solving the BCS problem based on Majorization Minimization (MM) technique. Our proposed approach is able to achieve significantly higher accuracy and shorter run times as compared to existing approaches.

## 1. INTRODUCTION
With an exponential rise in online resources and retailers the users are inundated with choices. The difficulty in evaluating huge number of options and selecting a few becomes a daunting task. This is what forms the ground for increased interest in design of efficient recommender systems (RS) in recent times. RS suggests a few items of relevance to the customers from the repository based on their past behaviour or choices. They make use of either explicit (ratings given by users) data or information extracted from implicit sources (browsing history or buying pattern) to make appropriate suggestions. Collaborative filtering (CF) [1] is the most popular approach for design of RS. These techniques basically rest on the assumption that if two users rate a few items similarly, they will exhibit similar choice for other items as well. This belief can be exploited in two ways - memory based approach and latent factor model. Memory based models [2] [3] are heuristic methods which use the available rating data to find users similar to the target user and a weighted average of ratings by similar users is used as to predict rating for the target user. On the other hand (latent factor) model based techniques [4] fundamentally represent the available information in the form of a lower dimensional structure under the proposition that the overall rating matrix can be represented by a relatively small number of independent variables or factors (Latent factor model). In latent factor models a user is modelled as a vector describing his/her affinity to the all the latent factors and items are modelled as vectors defined by the degree to which they possess each of the latent factor. They have been shown to outperform their memory based counterparts, in terms of Quality of Prediction (QoP) and coverage [5]. Most of the current research is centred on latent factor models. They mostly focus on improving the prediction accuracy by incorporating additional constraints or information into the existing framework. However, the basic model remains the same; involving factorization of rating matrix into two matrices one representing users and other items (1) - both being dense (dense solution promoted by Frobenius norm constraint).

$$\min_{U,V} \ \|Y - A(UV)\|_2^2 + \lambda \left( \|U\|_F^2 + \|V\|_F^2 \right) \quad (1)$$

where, $U$ and $V$ represent the user and item latent factor matrix respectively, $\lambda$ is the regularization parameter, $Y$ is the available set of ratings and $A$ is a linear operator. Traditionally two methods Stochastic Gradient Descent (SGD) (http://sifter.org/ simon/ journal/20061211.html) and Alternating Least Squares (ALS) [6] are commonly employed to solve the above problem.

Our work focusses on changing the basic structure of the latent factor model itself while retaining the primary concept that a limited number of (latent) factors affect the overall rating structure. We propose to factor the rating matrix into a dense user latent factor matrix and a sparse item factor matrix, by replacing Frobenius norm constraint on V by a sparsity promoting $l_1$ constraint (2).

$$\min_{U,V} \ \|Y - A(UV)\|_2^2 + \lambda_u \|U\|_F^2 + \lambda_v \|vec(V)\|_1 \quad (2)$$

where, $\lambda_u$ and $\lambda_v$ are the regularization parameters.

This stems from the reasoning that no item will ever possess all the factors and hence its latent factor vector will have zeros for attributes not possessed by the items. The remaining paper is organized as follows. In section 2, we discuss our problem formulation and proposed algorithm. Section 3 contains the experimental setup, results and comparisons with existing CF techniques. Paper ends with conclusion and future direction in section 4.

## 2. PROPOSED FORMULATION AND ALGORITHM
In this section, we discuss our novel proposition for latent factor model based formulation for design of an effectual recommender system.

## 2.1 Problem Formulation

Consider a rating matrix $\Re \in R^{M \times N}$ in which a value $\Re_{m,n}$ indicates the rating (explicit) given by $m^{th}$ user (out of $M$ total users) on the $n^{th}$ item (out of $N$ total items). As previously discussed, in latent factor approach each user/item can be modelled as a vector defining its affinity/degree of possession of a latent factor. For example, a book may be characterized in terms of features such as drama, comic, and sci-fi and a user vector will have values based on his liking/disliking of these features. Preference of a user for a particular item can hence be formulated as the interaction between their individual latent factor vectors as in (3)

$$\Re_{m,n} = \langle U_m, V_n \rangle \tag{3}$$

where, $V_n$ is the latent factor vector of item $n$ and $U_m$ is the $m^{th}$ user's latent factor vector.

However, the actual (explicit) ratings are not just a result of this interaction. They are plagued by certain inherent biases of users and items. For example, a critical user, giving lesser valued ratings, has an inherent negative bias, whereas a popular or award winning book/movie will tend to be given higher ratings by all users generally, thereby afflicting it with a positive bias. Thus, the actual rating can be modelled by modifying (3) by incorporating the bias terms (or baseline estimates) with the interaction part.

$$\Re_{m,n} = \mu_g + b_m + b_n + \langle U_m, V_n \rangle \tag{4}$$

where, $\mu_g$ is the global mean, $b_m$ is the bias of $m^{th}$ user and $b_n$ is the bias of $n^{th}$ item. $\mu_g + b_m + b_n$ forms the baseline part.

In our formulation, we estimate the baseline offline from the available dataset using method outlined in [7] by solving the following convex optimization framework.

$$\min_{b^*} \left\| \Re_{m,n} - b_m - b_n - \mu_g \right\|_2^2 + \delta \left( b_m^2 + b_n^2 \right) \tag{5}$$

where, $\delta$ is the regularization parameter. Once baseline is estimated, we extract the interaction part from the available ratings as $Z_{m,n} = \Re_{m,n} - \mu_g - b_m - b_n$.

Traditionally, Z is completely recovered from its sub sampled observation using matrix factorization approach outlined in (1). Equation (1) aims to recover the rating (interaction) matrix assuming that both user and item latent factor matrices have a dense structure or none of the terms in the latent factor vectors are zero for either users or items.

However, such an assumption is not entirely correct. Consider a RS which recommends movies to users. Each user can be described as a vector of values indicating his preference for a particular trait such as violence, musical, rom-com etc. Usually all users will have a certain affinity either for or against each of these traits. Thus the latent factor vector for any user will be dense, having non zero values throughout. However, the same reasoning cannot be applied to the items. Each item, say a movie in above scenario will either possess a trait or not. For example, a movie with a comic storyline or a children movie, will not have any presence of violence in it. Each item will thus be profiled by a latent factor vector which is sparse. There will be zeros in positions corresponding to traits which an item does not possess altogether. Thus the item matrix (V) will have sparse structure with several zeros in each column unlike the assumption made in previous models.

Carrying forward this logic, we put forth a new formulation for matrix factorization model. We aim to predict the missing ratings by formulating an optimization problem which promotes recovery of a dense user factor matrix (U) and a sparse item factor matrix (V) as in (2), repeated here for convenience.

$$\min_{U,V} \left\| Y - A(U \times V) \right\|_2^2 + \lambda_u \|U\|_F^2 + \lambda_v \|vec(V)\|_1 \tag{6}$$

Our formulation follows the Blind Compressive Sensing framework [8] [9] given in (7).

$$\min_{c, \hat{\Psi}} \left[ \sum_{i=1}^{l} \left\| M_i(c\hat{\Psi}) - y_i \right\|_2^2 \right] + \lambda \|c\|_1 \quad s.t. \quad \left\| \hat{\Psi} \right\|_F^2 \leq const \tag{7}$$

where, the task is to estimate the sparse representation (c) and sparsifying basis ($\hat{\Psi}$) from a given set of observations (y). M is the sub sampling operator and $\lambda$ is the regularization parameter. Compressive sensing framework focusses on estimating the sparse representation of a signal assuming that the sparsifiying basis is known a priori. BCS proposes an augmentation of CS theory, which assumes that the sparsifying basis is not known and the same is estimated along with the sparse coefficients only. The constraint on (bounded Frobenius norm) along with $l_1$ penalty term (on c) ensures uniqueness of the solution. Our formulation is similar to unconstrained form of (7) with the sparse coefficient set (c) equivalent to (V) and the dictionary being equivalent to (U).

Before deriving the algorithm for solving our problem, we will briefly justify the choice of BCS for solving our problem. According to the CS convention, $A$ is the sensing matrix and $U$ is the sparsifying basis / dictionary. For our problem $A$ is a Dirac / sampling operator; $A$ is the canonical basis. In order to satisfy the incoherence criterion demanded by CS the dictionary $U$ should be incoherent with $A$, i.e. it should be incoherent with the canonical basis. In other words, the elements in $U$ should be of small magnitude in order to ensure the coherence to be small. As the columns of $A$ are sparse (canonical basis), the rows of $U$ should be dense, i.e. should have small valued coefficients throughout. The $l_2$ penalty on the dictionary ($U$) enforces a minimum energy solution; it is well known that the minimum energy solution is dense with small values everywhere. This guarantees that the dictionary is incoherent with the sensing matrix $A$. Therefore, by design the BCS formulation yields a theoretically sound recovery framework for our problem.

## 2.2 Algorithm Design

To efficiently solve our formulation and recover the complete rating matrix, we propose an algorithm following the Majorization Minimization (MM) approach [11]. The use of MM approach enables us to break complex optimization problem into simpler and more efficiently solvable steps.

First we briefly discuss the MM approach to depict its advantage in reducing computational complexity. Consider an underdetermined system of equation, $y = Ax$ where, $A \in R^{l \times p}$, where $p > l$ is a fat matrix. As $A$ is not full rank, we can compute $x$ by least square minimization as $x = \left( A^T A \right)^{-1} A^T y$ which involves computation of inverse of large matrices. For cases where the variables to be recovered are very large, such as in recommender systems, the size of $A^T A$ becomes prohibitively large to efficiently compute its inverse within reasonable resource requirements. Use of MM approach eliminates the need to compute such inverses and reduces the computational burden significantly.

MM approach essentially involves replacing the minimization of an existing function, $F(x)$, by minimization of another function, $G(x)$, which is easier to minimize. $G(x)$ should be majorizer of $F(x)$ i.e.

1. $G(x) \geq F(x) \forall x$
2. $G(x) = F(x)$ at $x = x_k$

For the function $F(x) = \|y - Ax\|_2^2$, we can take $G(x)$ s.t. at $k^{th}$ iteration it is given by (8)

$$G_k(x) = \|y - Ax\|_2^2 + (x - x_k)^T (\beta I - A^T A)(x - x_k) \quad (8)$$

where, $\beta \geq \max eig(A^T A)$.

After some mathematical manipulation it can be shown that, $G(x)$ can be minimized iteratively using following steps.

$$z = x_k + \frac{1}{\beta} A^T (y - Ax_k) \quad (9)$$

$$\min_z \|z - x\|_2^2 \quad (10)$$

Now, we consider our problem formulation (6). We use ADMM (Alternate Direction Method of Multipliers), as the variables $U$ and $V$ are separable, to split (6) into 2 simpler sub problems.

Sub problem 1

$$\min_U \|Y - A(UV)\|_2^2 + \lambda_u \|U\|_F^2 \quad (11)$$

Sub problem 2

$$\min_V \|Y - A(UV)\|_2^2 + \lambda_v \|vec(V)\|_1 \quad (12)$$

We use majorization minimization, as described before, for solving both the sub problems.

Firstly, considering sub problem 1 (11) we can solve it in the following algorithmic steps using MM approach.

$$Z = U_k V_k + \frac{1}{\beta} A^T (Y - A(U_k V_k)) \quad (13)$$

Where, $\beta = \max eig(A^T A)$

$$\min_U \|Z - UV\|_2^2 + \lambda_u \|U\|_F^2 \quad (14)$$

Equation (14) of the above algorithm can be recast in following form

$$U \times N = D \text{ where, } D = VV^T + \lambda_u I \text{ and } N = ZV^T \quad (15)$$

*Initialization*: $U_0, V_0 = rand$
*Set regularization parameters*;
*Set maximum no. of iterations, N*
*while* $k < N$ *or* $obj\_func(k) - obj\_func(k-1) \leq 1e-7$

$$Z = U_k V_k + \frac{1}{\beta} A^T (Y - A(U_k V_k))$$

$$U_{k+1} \leftarrow Solve \; U \times [ZV^T] = [VV^T + \lambda_u I]$$

$$W = U_{k+1} V_k + \frac{1}{\beta} A^T (Y - A(U_{k+1} V_k))$$

$$V_{k+1} \leftarrow Soft\left(V + \frac{1}{\alpha} U^T (W - UV), \frac{\lambda_v}{2\alpha}\right)$$

*end while*

Equation (15) is just a linear system of equation which can be solved iteratively using any gradient descent algorithm.

2nd sub problem can also be similarly reorganized to give

$$W = U_{k+1} V_k + \frac{1}{\beta} A^T (Y - A(U_{k+1} V)) \quad (16)$$

$$\min_V \|W - UV\|_2^2 + \lambda_v \|V\|_1 \quad (17)$$

For solving (17), we use iterative soft thresholding as follows

$$V \leftarrow soft\left(T, \frac{\lambda_v}{2\alpha}\right)$$

Where, $T = V + \frac{1}{\alpha} U^T (W - UV)$ (18)

and $soft(t, s) = sign(t) \max(0, |t| - s)$

and $\alpha = 1.01 \times \max \; eig(U^T U)$

Both the sub problems are alternately solved till a desired number of iterations are complete or objective function converges. The complete algorithm is outlined in fig. 1.

## 3. EXPERIMENTAL SETUP AND RSULTS

We conducted experiment on the movielens 100K and 1M datasets (http://grouplens.org/datasets/movielens/). The 100K dataset consist of 100,000 ratings (rating value ranging from 1-5) given by 943 users on 1682 movies. 1M dataset consists of ratings on 3952 movies by 6040 users. Both the datasets are divided into test and training data to perform cross validation. We conducted three, five and ten fold cross validation to vary the ratio of sizes of test and training data. For *'n'* fold cross validation the entire dataset is split into *n* blocks. *'n-1'* blocks are combined to form the training data and the $n^{th}$ set is taken as the test set. The simulations are carried out on system with i7-3770S CPU @3.10GHz with 8GB RAM.

For our algorithm (BCS-CF), the value of regularization parameter, $\delta$ in (5) is taken to be *1e-3*. The value of regularization parameters for (6) is kept at $\lambda_u = 1e3$ and $\lambda_v = 1e-1$ for the 100K dataset and $\lambda_u = 1e4$ and $\lambda_v = 1e-1$ for the 1M dataset. The values are achieved using L-curve technique [12]. The number of latent factors (rank) of user/item matrices is taken to be 50. The values are selected after extensive experimentation.

The results of our formulation is compared against the results obtained using SGD formulation proposed in [7], Accelerated Proximal Gradient based matrix completion (APG) algorithm [13], Matrix completion using Split bregman (MCSB) [14] and optSpace [15]. We also compare our results against the BCS algorithm (BCSJ) put forth in [9] to show the efficiency of our algorithm. The algorithms are compared terms of Quality of prediction, measured in terms of Mean Absolute Error (MAE) (19) and on the basis of run times.

$$MAE = \frac{\sum_{m,n} \Re_{m,n} - \hat{\Re}_{m,n}}{|\Re|} \quad (19)$$

where, $\Re_{m,n}$ and $\hat{\Re}_{m,n}$ are the actual and predicted ratings and $|\Re|$ is the cardinality of the rating matrix $\Re$.

**Table 1. MAE for 100K dataset**

| Algorithm | MAE-3 Fold | MAE-5 Fold | MAE-10 Fold |
|---|---|---|---|
| BCS-CF | 0.7417 | 0.7215 | 0.7140 |
| MC-SB | 0.7454 | 0.7323 | 0.7279 |
| APG | 0.8573 | 0.8847 | 0.9187 |
| OptSpace | 0.7629 | 0.7450 | 0.7323 |
| BCS-J | 0.7527 | 0.7430 | 0.7414 |
| SGD | 0.8002 | 0.7432 | 0.7312 |

**Table 1. MAE for 1M dataset**

| Algorithm | MAE-3 Fold | MAE-5 Fold | MAE-10 Fold |
|---|---|---|---|
| BCS-CF | 0.6835 | 0.6762 | 0.6712 |
| MC-SB | 0.6999 | 0.6943 | 0.6897 |
| APG | 1.0178 | 0.9782 | 0.9352 |
| OptSpace | 0.6907 | 0.6886 | 0.6844 |
| BCS-J | 0.6967 | 0.6917 | 0.6858 |
| SGD | 0.6988 | 0.6936 | 0.6907 |

Table 1 gives the MAE values of all the algorithms for 3, 5 and 10 fold cross validation. The results shown are averaged values for 100 independent runs of each algorithms on each test-train pair. It is evident from all the three sets of data that our algorithm perform better than both the other techniques compared against. Our algorithm gives an improvement in recovery accuracy of around 3% (for 10 fold validation) to 8 % (for 3 fold cross validation) over SGD. Thus, it is clear that performance of SGD deteriorates as the size of available training set reduces (10 fold to 3 fold) but our algorithm consistently performs well. As compared to other state of the art matrix completion approaches (APG, optSpace and MCSB) our algorithm gives an improvement of atleast 2%.

Table 2 gives the MAE values on 1M dataset for all the three methods. Even for the 1M dataset, our algorithm gives improved results over others compared against (around 3 % improvement in MAE).

Table 3 gives the average run times of all the algorithms for both 100K and 1M datasets for 5-fold validation. The run times for other cross validation schemes are also in close vicinity. From entries in table 3, it is clear that SGD is very slow (40 times slower) as compared to our formulation. This is mainly because of the large number of iterations involved in the former. The BCS-J algorithm also is slower than ours by around 5 times. Our formulation is around 4 times faster than the fastest of the algorithms compared against i.e. optSpace. As the size of dataset increases (100K to 1M), the time required also increases and thus our algorithm is better suited for large recommender systems viz-a-viz others. A faster algorithm is not only more appropriate for online computations, it also reduces the burden on the system if recalculations need to be made when a new user/item is added. Thus updation can be carried out more promptly.

**Table 3. Run Times–5 Fold**

| Algorithm | Run Times-seconds (100K) | Run Times-seconds (1M) |
|---|---|---|
| BCS-CF | 2.67 | 31.36 |
| MC-SB | 61.5 | 979.23 |
| APG | 15.01 | 228.5 |
| OptSpace | 8.65 | 175.23 |
| BCS-J | 18.74 | 153.21 |
| SGD | 150.34 | 1262.5 |

## 4. CONCLUSION

In this work, we propose a novel formulation based on latent factor model for collaborative filtering. The basic formulation remains the same, i.e. the ratings matrix can be factored into two matrices – user latent factor matrix and item latent factor matrix. However all prior studies assumed that both these matrices are dense. We argue that this is not the case; the user latent factor matrix is dense but the item latent factor matrix is supposed to be sparse. This is because all users are likely to have an affinity for all the different factors, but it is not possible for the items to possess all factors simultaneously.

We show that our proposed formulation naturally fits into the recently proposed blind compressive sensing framework. BCS is a recent framework and there are no efficient algorithms for solving the BCS problem. In this paper we propose a majorization minimization algorithm to solve the BCS problem. Use of MM approach greatly reduces the computational complexity and hence execution times of our algorithm. Our claim is braced by comparison of run times of other methods viz-a-viz ours. A fast algorithm aids in easier updation of model with addition of new users or items along with faster online computations.

In this work we have experimented on movie datasets. The proposed formulation yields significant improvement over existing matrix factorization models. In the future, we would like to see if our proposal enjoys the same success for other recommendation systems (books, garments etc.) as well.